\documentclass[prl]{revtex4}
\usepackage{bm}
\usepackage{amsmath}
\usepackage{amssymb}
\usepackage{amsxtra}
\usepackage{amscd}
\usepackage{amsthm}
\usepackage{amsfonts}
\usepackage{eucal}
\usepackage{latexsym,amsthm}
\usepackage{pstricks}
\usepackage{color}
\usepackage{graphicx}

\usepackage{natbib}

\usepackage{setspace}

\newcommand{\nc}{\newcommand}
\nc{\on}{\operatorname}
\nc{\wt}{\widetilde}
\nc{\Wick}{{\mathbb :}}
\nc{\R}{{\mathbb R}}

\newcommand{\beq}{\begin{equation}}
\newcommand{\eeq}{\end{equation}}
\newcommand{\bmul}{\begin{multline}}
\newcommand{\emul}{{\end{multline}}}
\newcommand\beqa{\begin{eqnarray}}
\newcommand\eeqa{\end{eqnarray}}
\newcommand\bea{\begin{array}}
\newcommand\eea{\end{array}}
\newcommand\ba{\begin{array}}
\newcommand\ea{\end{array}}

\newcommand{\neqa}{\nonumber\end{eqnarray}}

\newcommand{\eq}[1]{eq.(\ref{#1})}

\newcommand{\Eq}[1]{Eq.(\ref{#1})}

\newcommand{\Tr}{{\rm Tr}}

\newcommand{\<}{{\langle}}
\renewcommand{\>}{{\rangle}}

\nc{\CH}{{\mathcal H}}
\nc{\Db}{{\bar D}}
\nc\comment[1]{}

\nc{\CM}{{\mathcal M}}
\nc{\CN}{{\mathcal N}}

\newcommand{\bk}{{\bf k}}
\newcommand{\bp}{{\bf p}}
\newcommand{\bq}{{\bf q}}

\newcommand{\re}{\relax{\rm I\kern-.18em R}}

\newcommand{\mean}[1]{\left\langle #1\right\rangle}
\newcommand{\abs}[1]{\vert #1\vert}

\newcommand{\ket}[1]{\vert #1 \rangle}
\newcommand{\bra}[1]{\langle #1\vert}

\nc{\meV}{{\mathrm{\,meV}}}
\nc{\cG}{{\mathcal G}}

\renewcommand{\)}{\right)}
\renewcommand{\(}{\left(}
\renewcommand{\bar}{\overline} 

\nc{\al}{{\alpha}}

\def\eps{{\epsilon}}
\begin{document}

\title{Nematic phase in a two-dimensional Hubbard model at weak coupling and finite temperature}
\author{
Sergey Slizovskiy $^{1,2,6}$, Pablo Rodriguez-Lopez $^{1,3,4,5}$ and Joseph J. Betouras $^{1}$} 
\affiliation{
$^{1}$ Department of Physics and Centre for the Science of Materials,  Loughborough University, LE11 3TU, UK \\
$^{2}$ National Graphene Institute, The University of Manchester, M13 9PL, Booth st. E.,  Manchester, UK \\
$^{3}$ Materials Science Factory, Instituto de Ciencia de Materiales de Madrid, ICMM-CSIC, Cantoblanco, E-28049 Madrid, Spain \\
$^{4}$ Department of Physics, University of South Florida, Tampa FL, 33620, USA \\
$^{5}$ GISC-Grupo Interdisciplinar de Sistemas Complejos, 28040 Madrid, Spain \\
$^{6}$ on leave from: NRC ``Kurchatov Institute'' PNPI,  Gatchina, 188300, Russia  
}

\begin{abstract}
We apply the self-consistent renormalized perturbation theory to the Hubbard model on the square lattice, 
at finite temperatures in order to study the evolution of the Fermi-surface (FS)  as a function of temperature and doping.  Previously, a nematic phase for the same model has been reported to appear at weak coupling near a Lifshitz transition from closed to open FS at zero temperature where the self-consistent renormalized perturbation theory was shown to be sensitive to small deformations of the FS. We find that  the competition with the superconducting order leads to a maximal nematic order appearing at non-zero temperature.  We explicitly observe  the two competing phases near the onset of nematic instability and, by comparing the grand canonical potentials, we  find that the transitions are first-order.   We explain the origin of the interaction-driven  spontaneous symmetry breaking to a nematic phase in a system with several symmetry-related Van Hove points and discuss the required conditions. 
\end{abstract}
\maketitle

\section{Introduction}
The two dimensional Hubbard model (HM), one of the most fundamental and widely-used models in condensed matter physics, still presents a major theoretical challenge. Many different modern techniques have been used to provide converging results \cite{Leblanc_etal_2015}. Recently the ground-state phase diagram of the repulsive HM with nearest and next nearest hoping elements from the weak-coupling point of view revealed a very rich behavior \cite{Deng_EPL_2015, Kozik16} in a wide range of parameters, extending earlier works \cite{Hlubina, Raghu_Kivelson_Scalapino_2010}. A great variety of techniques has been used, ranging from random phase approximation and Gutzwiller approximation to renormalization group theory and parquet diagrams and focusing on the interplay between magnetic and superconducting phases with the aim to account for the strong correlations \cite{Kotliar, Irkhin_Katanin_Katsnelson}. 

One of the phases that is considered as the preferred ground state in a range of parameters, is the nematic phase which can be the ground state of a strongly correlated system under certain conditions  \cite{Kivelson_Fradkin_Emery, Oganesyan_Kivelson_Fradkin, Metzner1, Metzner2, Kivelson_RMP_2003, Yamase_Metzner_2007, Kee_etal_2003, Kim_Kee_2004, Khavkine_etal_2004, Kitatani_etal_2017, Kaczmarczyk_etal_2016, Doan_Manousakis}. { Indeed, after the insightful proposal that correlated electron systems, seen as electronic fluids, can host different phases in a direct analogy with classical fluids with different degrees of translational and rotational symmetry breaking \cite{Kivelson_Fradkin_Emery}, a plethora of studies and models appear that exhibit a nematic phase as a ground state in a certain parameter range. In an electronic nematic phase the rotational invariance is broken in real space or even in both real and spin spaces, leading to more exotic ground states.}

From the experimental point of view, general Fermi surface (FS) deformations, either non-topological, such as various Pomeranchuk instabilities associated with a symmetry-breaking \cite{Pomeranchuk, HalbothMetzner, Quintanilla_Schofield, Fradkin_etal_2010}  or topological with no symmetry breaking (Lifshitz transitions) \cite{Lifshitz, SSJoseph, Carr_Quintanilla_Betouras, Ghamari_etal, Hackl_Vojta, Goldstein_PRB_2017, Liu_etal, Yelland_etal, Varlet_etal, Slizovskiy_Chubukov_Betouras, Shtyk_Goldstein_Chamon}, play a very important role due to an enhanced or even singular density of states and novel orders that are associated with them. Especially the appearance of a nematic phase in systems like cuprates and pnictides \cite{Fradkin_etal_2010, Hinkov_etal_Science_2008, Lawler_etal_Nature_2010, Daou_et_al_Nature_2010, Ramshaw_etal_2017, Fernandes_Chubukov_Schmalian_2014, Watson_etal_2015} posed many fundamental questions. It is therefore important to understand a basic model like the HM in different parameter ranges so that a better physical picture can be formed and puzzling experimental results can be explained \cite{Slizovskiy_Chubukov_Betouras, Hackl_Vojta}.

{ The current numerical approaches for strongly correlated electron systems are in general not well suited to capture the effects of FS deformations either within a nematic phase or in the case of competing phases. Most of the problems stem from the fact that the systems are finite size (for example dynamical mean-field theory or dynamic cluster approximation or variational Monte Carlo). Recently a variational method based on Gutzwiller wave function combined with a diagrammatic expansion technique was applied to the HM  \cite{Kaczmarczyk_etal_2016} with the main result that the coexistence of nematic (breaking of the four-fold $C_4$ symmetry) and superconducting order turn a d-wave order parameter to d+s as expected (which in high-T$_c$ superconductors happen anyway due to orthorhombic distortion \cite{Betouras_1995}). Even more recent studies using the fluctuation exchange approximation combined with dynamical mean-field theory led to the same conclusion \cite{Kitatani_etal_2017}. As a general conclusion,} these numerical approaches are either limited to rather high temperatures but do not fully account for long-range interactions or are very powerful, such as the diagrammatic Monte Carlo technique, and are based on perturbation theory to sum up the relevant diagrams.  Therefore, the physical insight that can be developed by perturbation theory is quite important.  

In this work, we focus on the physics in the vicinity of the Lifshitz transition of the HM, when the nematic state can be developed. We compute the self-energy, the nematic and superconducting order parameters and characterise the order of the transition. We discuss in a heuristic way the conditions under which the breaking of the $C_4$ symmetry occurs. A tractable and relatively computationally inexpensive approach, the renormalized perturbation theory
was formulated in Ref.[\onlinecite{Metzner1}] following earlier ideas \cite{Nozieres, Salmhofer}.  
An advantage of this method is that all possible orderings of the system, with their accompanied order parameters, will appear automatically in the geometry of the FS and in the values of the gap function and need not be introduced by hand.  
This makes this 
formalism very appealing for studies of various Pomeranchuk instabilities,  with the nematic instability as the prominent example. 

In the following sections,  first we review for completeness and generalize to finite temperatures the formalism developed in Ref. \cite{Metzner1}. Subsequently, we apply this method to the 2D HM, with the same parameters to seek agreement at $T=0$. 
Our results confirm the presence of nematic phase in the vicinity of the topological Lifshitz transition of the 2D FS. We elaborate on the details of the nematic phase transition and find it 
to be of the first order. Effectively, the topological Lifshitz transition, which had been shown to be of the first order\cite{SSJoseph} when there is another FS as a particle reservoir,  is split by the appearing nematic phase into two 
first-order phase transitions.  The generalization to finite temperature allows us to study the competition of the nematic and superconducting instabilities, where we find that the nematic order survives to  higher
temperatures compared to superconductivity and the nematic instability  is the strongest at non-zero temperatures, when superconductivity starts to be suppressed. 
Computation of the free energy allows the comparison of the coexisting phases and the identification of the locations of first-order phase transitions.

\section{Finite temperature self-consistent perturbation theory}

In this method, the self-energy at the Fermi-surface (FS), which is the main quantity of interest, is built into the renormalized Green's function. The general Hamiltonian of interacting fermions, written as a non-interacting $H_0$ and an interacting $H_I$ part $H= H_0 + H_{I} = \sum_{\bk,\sigma} \xi_{\bk} n_{\bk, \sigma} + H_I$, with $\xi_{\bk} = \eps_{\bk} - \mu$  is split as  $H = \tilde H_0 + \tilde H_I = (H_0 + \delta H_0) + (H_I - \delta H_0)$, with the quadratic counter-terms $\delta H_0$ chosen in such a way that $\tilde H_0$ does already provide the correct Fermi surface (FS)  and $\tilde H_I$ does not carry any other change of the FS. As a consequence, the divergencies due to self-energy insertions are eliminated since the poles of the bare quasiparticle Green's function  do not need to be moved.  
This is equivalent to the statement that self-energy of the theory with counter-terms must vanish
on the FS. 
\beq \label{Main Equation}
 \tilde \Sigma(0, \bk) = 0 \text{ for } \bk \in \text{Fermi Surface} 
\eeq

\noindent The FS position in the above equation depends on the self-energy $\tilde\Sigma$  which does depend on the 
position of the FS when computed within the renormalized perturbation theory.    \Eq{Main Equation} is usually solved iteratively by computing 
the self-energy and correcting the FS position.  
The counter-terms are only fixed on the FS, therefore there is an ambiguity of their definition at other 
momenta.  A possible simple choice is to divide the Brillouin zone into sectors, with each sector crossing the FS once, and extend the counter-terms from the Fermi-surface
to the sectors as constants (i.e. pull-back the 
counter-terms).  Evidently, summation of all orders of perturbation theory provides a result which is independent of the choice of the counter-terms. At the same time, the results of the calculations, truncated to a finite order of 
perturbation theory will depend on the choice of this construction, but it provides 
a very good estimate of the precision of the approximations. 

The 
procedure can be extended  to the case where a superconducting instability is present. In that case, the matrix self-energy in Nambu formalism is required to vanish 
on the Fermi surface (FS)\cite{Metzner1}. 
The position of the FS and the zero-frequency gap function at each point of the FS are determined self-consistently. In the remaining section,  the method is generalized to finite temperatures, in which case the FS is defined by the location in reciprocal space where 
the energy dispersion renormalized by the 
self-energy, analytically continued at zero frequency, vanishes.  
The shape of the FS is then studied  by discretizing it into a large number of sectors 
and self-consistently solving \Eq{Main Equation}.   

\begin{figure}
\begin{center}
\includegraphics[scale=0.35]{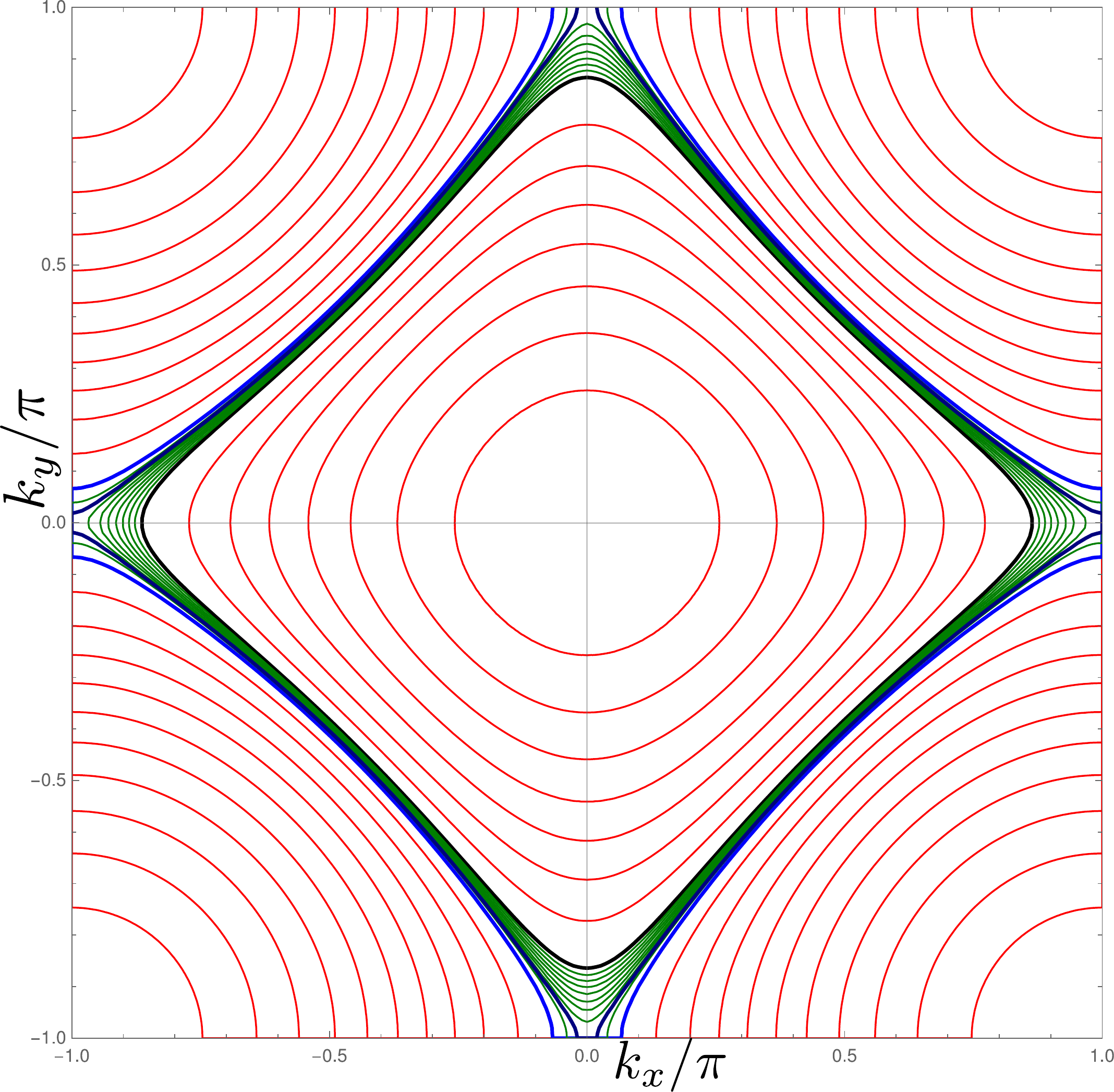}
\end{center}
\vspace{-0.7 cm}
\caption{\label{fig:FSIllustration} 
Fermi surfaces of t-t' Hubbard model on the square lattice, $t'=0.15 t$, $U=0$. Particle density varies from $n=0.1$ to $n=1.9$  with the range 0.8 to 0.9 highlighted. }
\end{figure}

The starting point of our work is the 2D HM:
\begin{eqnarray}
H = \sum_{\textbf{k},\sigma}\xi_{\textbf{k}}n_{\textbf{k},\sigma} + U\sum_{\textbf{k}}n_{\textbf{k},\downarrow}n_{-\textbf{k},\uparrow},
\end{eqnarray}
where
$\xi_{\textbf{k}} = \epsilon_{\textbf{k}} - \mu
\,$,
  $\mu$ is the chemical potential, $U$ is the interaction ($U>0$ is repulsive interaction) and the dispersion $\epsilon_{\textbf{k}}$ is chosen to represent the  first ($-t$) and second ($t'$) neighbor hopping tight-binding model on a square lattice:
\begin{eqnarray}
\epsilon_{\textbf{k}} = - 2t( \cos(a\,k_{x}) + \cos(a\,k_{y}) ) + 4t'( \cos(a\,k_{x})\cos(a\,k_{y}) - 1 )
\end{eqnarray}
The Brillouin zone is the square $[-\pi,\pi]\times[-\pi,\pi]$ (we set the lattice parameter $a=1$), and typical FS looks as sketched in Fig.\ref{fig:FSIllustration}.

At weak coupling,  the only expected instability for a symmetric dispersion under $k \to - k$  is a 
superconducting instability which can be included as a counter-term together with the self-energy at the Fermi surface:
\begin{eqnarray}
H = \sum_{\textbf{k},\sigma} \xi_{\textbf{k}} n_{\textbf{k},\sigma} + \left[\sum_{\textbf{k},\sigma} \delta \xi_{\textbf{k}} 
n_{\textbf{k},\sigma} + \sum_{\textbf{k}}\left( \Delta_{\textbf{k}}a^{\dagger}_{-\textbf{k}\downarrow}a^{\dagger}_{\textbf{k}\uparrow} + 
\Delta^{*}_{\textbf{k}}a_{\textbf{k}\uparrow}a_{-\textbf{k}\downarrow} \right) \right ] + 
U\sum_{\textbf{k}}n_{\textbf{k},\downarrow}n_{-\textbf{k},\uparrow} - \left[ ... \right],
\end{eqnarray}
where the counter-terms in square brackets comprise the gap function $\Delta_{\textbf{k}}$  as well as $\delta\xi_{\textbf{k}}$. 
In terms of Nambu operators,
\begin{eqnarray}
\Psi_{\textbf{k}} = \left(\begin{array}{c}
a_{\textbf{k}\uparrow}\\
a^{\dagger}_{-\textbf{k}\downarrow}
\end{array}\right)
\hspace{1cm}
\text{and}
\hspace{1cm}
\Psi^{\dagger}_{\textbf{k}} = \left(
a^{\dagger}_{\textbf{k}\uparrow},
a_{-\textbf{k}\downarrow}
\right),
\end{eqnarray}
Defining the renormalized dispersion, $\tilde\xi_\bk =\xi_\bk + \delta \xi_\bk$,  the renormalized quadratic part of the Hamiltonian becomes  
\begin{eqnarray}\label{Nambu_H0}
H_{0} = \sum_{\textbf{k},\sigma}\tilde\xi_{\textbf{k}}\Psi^{\dagger}_{\textbf{k}}\sigma_{3}\Psi_{\textbf{k}} - 
\sum_{\textbf{k}}\Psi^{\dagger}_{\textbf{k}}\left[ \Delta'_{\textbf{k}}\sigma_{1} - 
\Delta''_{\textbf{k}}\sigma_{2}\right]\Psi_{\textbf{k}},
\end{eqnarray}
where $\Delta_{\textbf{k}} = \Delta'_{\textbf{k}} + i\,\Delta''_{\textbf{k}}$, and $\sigma_{i}$ are the Pauli matrices.
The corresponding Nambu matrix propagator $\mathbb{G}_{0}(k) = \mean{\Psi\Psi^{\dagger}}_{0}$ is obtained as
\begin{eqnarray}
\mathbb{G}_{0}^{-1}(k) & = & \left(\begin{array}{cc}
i\omega - \tilde{\xi}_{\textbf{k}} & \Delta_{\textbf{k}}\\
\Delta^{*}_{\textbf{k}} &   i\omega + \tilde{\xi}_{-\textbf{k}}
\end{array}\right),
\end{eqnarray}
using $\xi(\bk) = \xi(-\bk)$ we have
\begin{eqnarray}
\mathbb{G}_{0}(k) & = & \left(\begin{array}{cc}
G_{0}(k) & F_{0}(k)\\
F^{*}_{0}(k) & -G_{0}(-k)
\end{array}\right)
 = \frac{1}{\omega^{2} + \tilde{\xi}^{2}_{\textbf{k}} + \abs{\Delta_{\textbf{k}}}^{2}}\left(\begin{array}{cc}
- i\omega - \tilde{\xi}_{\textbf{k}} & \Delta_{\textbf{k}}\\
\Delta^{*}_{\textbf{k}} &  i\omega - \tilde{\xi}_{-\textbf{k}}
\end{array}\right).
\end{eqnarray}
The matrix self-energy is defined as
\beq
\bf{\Sigma}(\textbf{k}) = \left(\begin{array}{cc}
\Sigma(\textbf{k}) & S(\textbf{k})\\
S^{*}(\textbf{k}) & - \Sigma(-\textbf{k})
\end{array}\right) ,
\eeq
and the  diagrams up to the second order give:
\beqa \label{SigmaEq}
\Sigma(k) &=& -\delta\xi + U \int_p G_0(p) +  U^2 \int_q G_0(k-q) \Pi(q), \\
S(k) &=& -\Delta_\bk - U \int_p F_0(p) - U^2 \int_q F_0(k-q) \Pi(q),
\label{SEq}
\eeqa

\noindent where we use a standard summation notation for Matsubara frequencies, for fermions $\int_{p}f(p) = \int_{p}f(\omega, \textbf{p}) \equiv  T\sum_{n\in\mathbb{Z}}\int\frac{d^{2}\textbf{p}}{(2\pi)^{2}}f(i\pi(2n + 1)T, \textbf{p})$
and bosons $\int_{q}b(q) = \int_{q}b(\omega, \textbf{q}) \equiv T\sum_{n\in\mathbb{Z}}\int\frac{d^{2}\textbf{q}}{(2\pi)^{2}}b(i2\pi n T, \textbf{q})$.

The polarization $\Pi(q)$ is defined as
\beq
 \Pi(q) = - \int_p \left[ G_0(p) G_0(p+q) + F_0(p) F_0^*(p+q) \right], 
\eeq
All the  frequency summations in the above formulae are evaluated analytically, this is presented in a separate file as Supplemental Material and an example of the results is: 
\beq  \label{S1term}
  U \int_p F_0(p) =  U \int \frac{d^2\bp}{(2 \pi)^2} \Delta(\bp) \( \frac{1}{2 E_\bp} - \frac{n_F(E_\bp)}{E_\bp} \)
\eeq
where $E_\bk \equiv \sqrt{\tilde \xi_\bk^2 + |\Delta_\bk|^2}$.
  Setting the frequency $\omega$ to zero leads to a natural finite-temperature generalization of self-consistent perturbation theory  \Eq{Main Equation},  where the counter-terms 
$\delta\xi_\bk$ and $\Delta_\bk$, defined on the FS, are calculated as the matrix self-energy is vanished on the FS:
\beq  \label{Consistency}
\Sigma(i 0, \bk) = 0  \text{ and } S(i 0, \bk) = 0    \text{ when  } \xi_\bk + \delta\xi_\bk = 0. 
\eeq 
$\Sigma$ and $S$ are solved iteratively by evaluating the $U^2$ terms for $\Delta(\bk)$, while the FS is calculated at the previous iteration 
and then both $\Delta$ and the FS are updated.  The chemical potential can be also tuned at each iteration so that the particle density is kept fixed \cite{Metzner1}, 
\beq
n = 2 \int_k G_0(k) = 2  \int \frac{d^2\bk}{(2 \pi)^2} \( \frac{E_\bk - \tilde \xi_\bk}{2 E_\bk} + \frac{n_F(E_\bk) 
\tilde\xi_\bk}{E_\bk} \) , 
\eeq

In the present work, the chemical potential is kept fixed so that several coexisting phases near the first order phase transition can be observed.  
To find $S$ at each iteration the first order self-consistency equation is solved with a second-order assumed fixed by rewriting:
Eqs.(\ref{SEq}, \ref{S1term}, \ref{Consistency}) as
\beq
0 = -\Delta_\bk - U \int  \frac{d^2\bp}{(2 \pi)^2} \Delta(\bp) \( \frac{1}{2 E_\bp} - \frac{n_F(E_\bp)}{E_\bp} \) - U^2 \int_q 
F_0(k-q) \Pi(q)
\eeq
To solve it for $\Delta(k)$ it is useful to note that the entire momentum dependence of $\Delta(k)$ is coming from the $U^2$ term.
Therefore, choosing an arbitrary point $\tilde \bk$ on the Fermi-surface, and denoting the second-order term as 
\beq
S^{(2)}(\bk) = - U^2 \left.\int_q F_0(k-q)  \Pi(q) \right|_{\omega \to 0 i},
\eeq
 the momentum dependence of the gap function is obtained:
\beq
  \Delta(\bk) = \Delta(\tilde\bk) +  S^{(2)}(\bk) - S^{(2)}(\tilde \bk)= \Delta(\tilde\bk) + \delta\Delta(\bk),
\eeq 
and the $\Delta(\tilde\bk)$ is extracted:
\beq
  \Delta(\tilde \bk) =  \frac{- U \int \frac{d^2\bp}{(2 \pi)^2} \delta\Delta(\bp) \( \frac{1}{2 E_\bp} -  
\frac{n_F(E_\bp)}{E_\bp} \) + S^{(2)}(\tilde\bk)}{ 1 + U \int \frac{d^2\bp}{(2 \pi)^2} \( \frac{1}{2 E_\bp} -  
\frac{n_F(E_\bp)}{E_\bp} \) }.
\eeq 
The iterations are repeated until convergence is achieved. Notably, ordinary iterations do not normally converge near the phase transition point (in a narrow region of shallow dispersion near the Van Hove point,
large and strongly non-linear change of Fermi surface arises in response to a small variation of self-energy). Thus, to achieve convergence,
the iteration step is decreased as $\xi(k) ^{(n+1)} = c\, \xi \left [ \xi(k)^{ (n) }, \Delta(k)^{(n)}
\right]  + (1 - c)\, \xi(k)^{ (n)} $ 
with a chosen $c<1$. 


To study the nematic phase transition, the grand canonical potentials $\Omega = F - \mu n $ of competing phases at equal chemical potentials are compared. 
The grand canonical free energy $\Omega$, is expressed  \cite{Metzner1} in the second order of  the renormalized perturbation theory  as 
\beq
 \Omega = \Omega_0 - \frac12 U^2 \int_q \Pi(q)^2, 
\eeq
where the free part is :
\beq
\Omega_0 =  - \int \frac{d^2 k}{(2 \pi)^2} \left[(E_{\bf k} -\xi_{\bf k})+ 2 T \log \(1 + e^{- E_{\bf k}/T } \) \right].
\eeq

The  factor $-\xi_{\bf k}$ in $\Omega_0$ can be traced back to an extra fermion operator ordering term, that appears when switching to Nambu formalism. For completeness, the derivation is presented in the Appendix.



\section{Results}
\begin{figure}
\begin{center}
\includegraphics[scale=0.4]{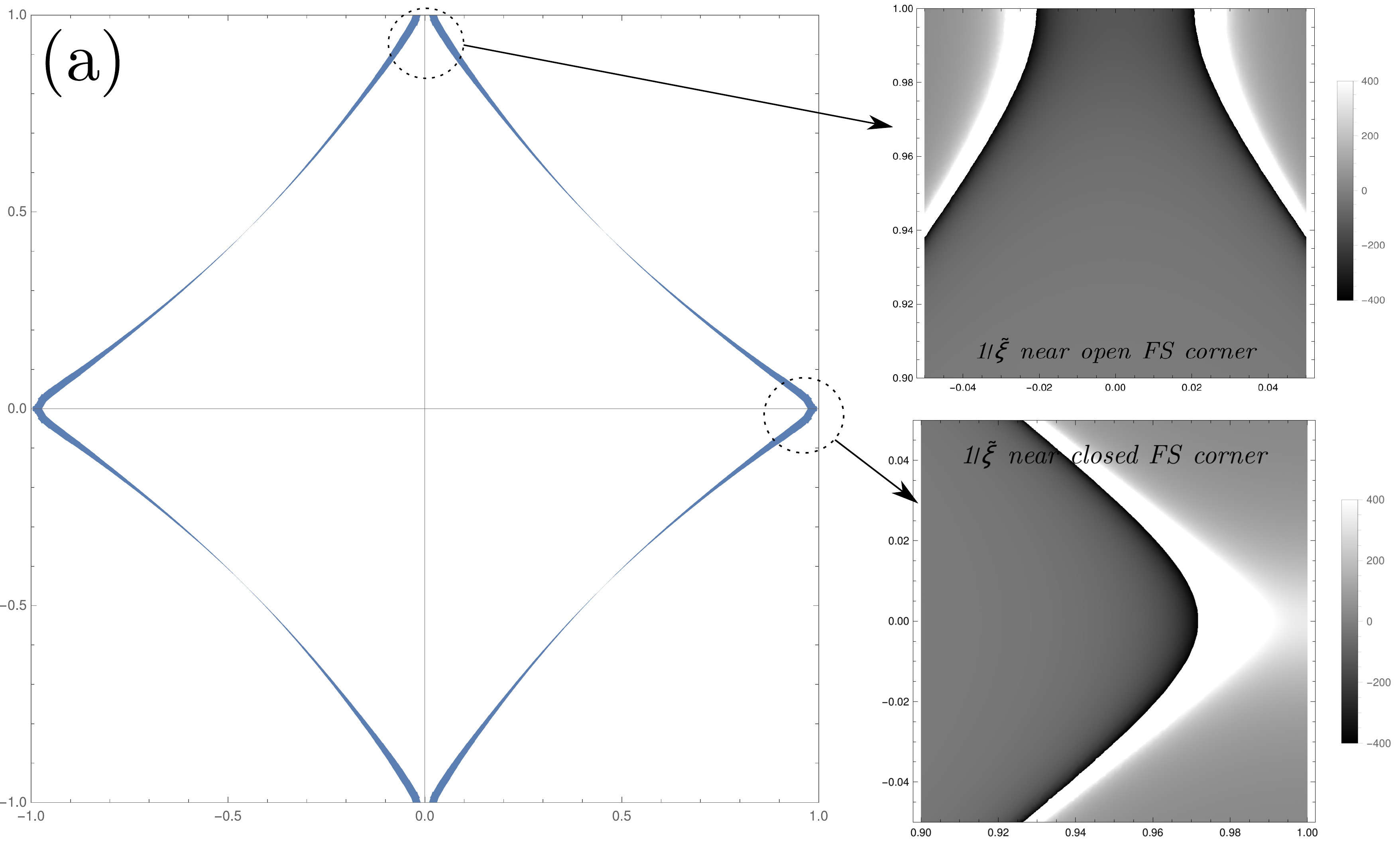}
\includegraphics[scale=0.5]{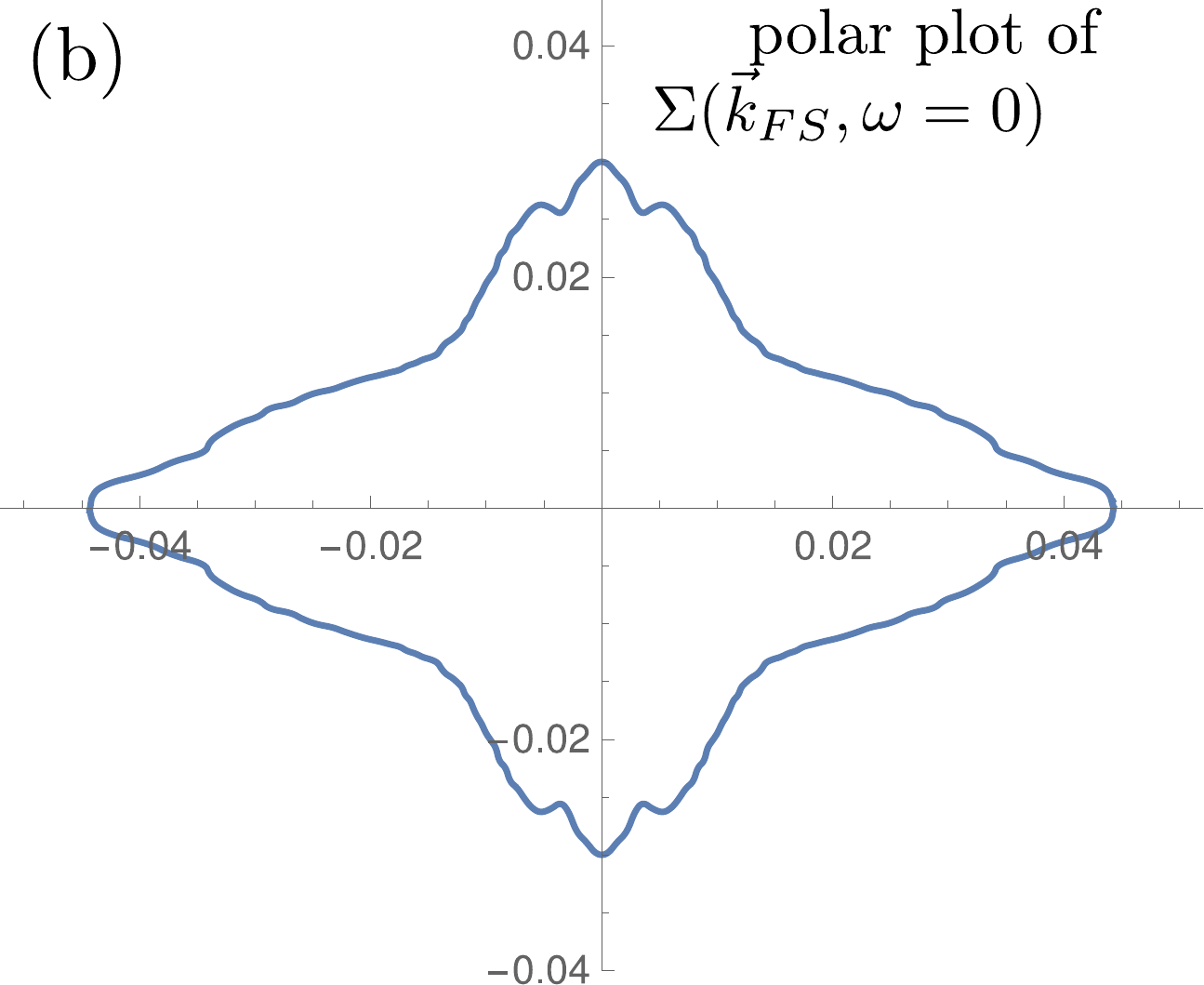}
\includegraphics[scale=0.4]{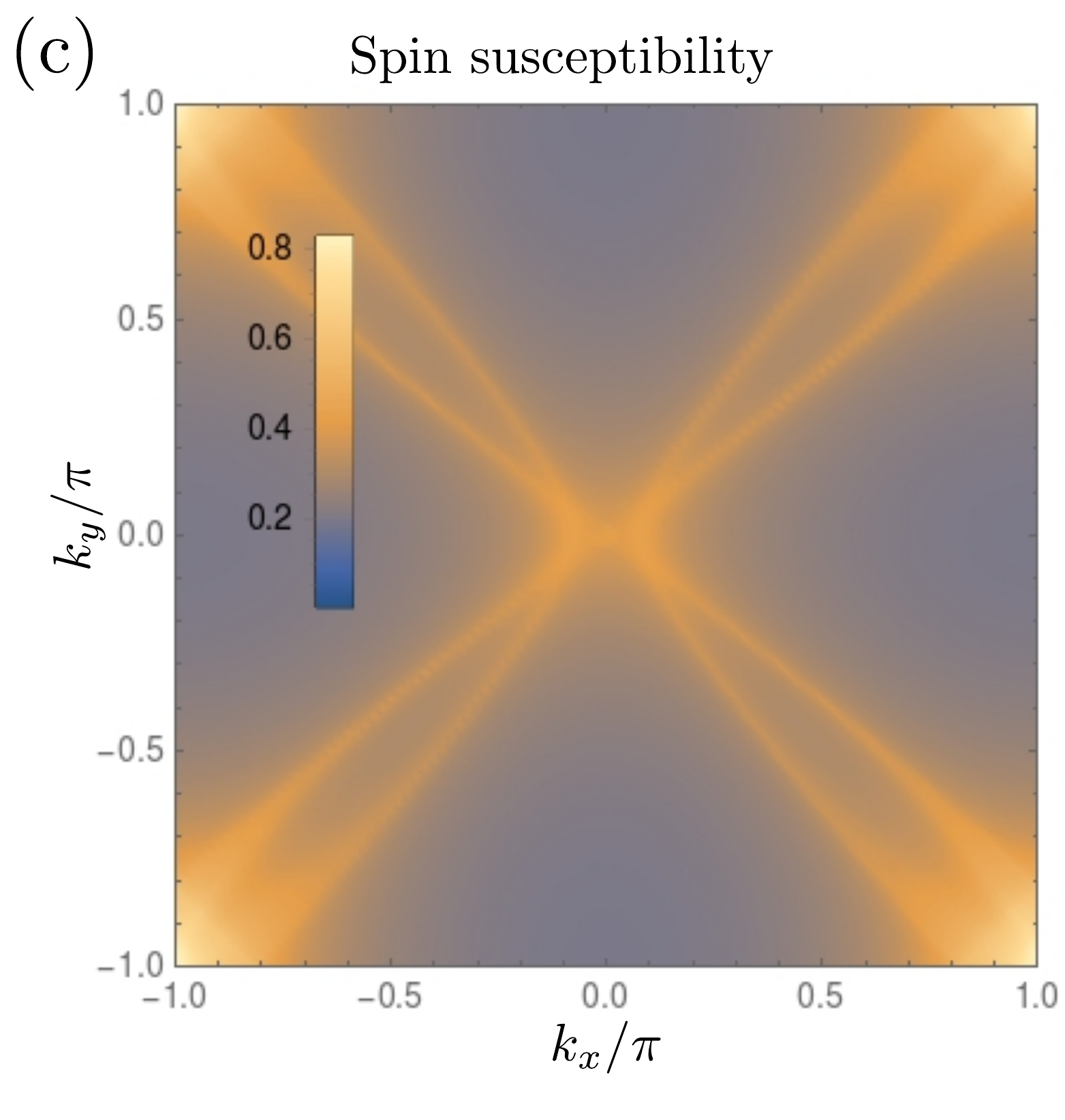}
\end{center}
\vspace{-0.7 cm}
\caption{\label{fig:NematicExample} 
(a) Typical nematic Fermi-surface for $n=0.883$ and $T = 0.001 t$ that one gets from iterations near the Lifshits transition.  
Line is bounded by $\tilde\xi(\bk) \pm \Delta(\bk) = 0$ contours, i.e. line width reflects the superconducting 
gap.  Density plots illustrate that $1/\tilde \xi$ is dominated by a positive/negative  contribution near the closed/open corners of the FS.  
(b) Polar plot of self-energy at the Fermi-surface in the  nematic phase (a constant has been added to make it positive, $n=0.884, T=0.001t$)  (c) Example of spin susceptibility $-U \Pi(0,\bq)$ in the nematic phase.   
}
\end{figure}

\begin{figure}
\begin{center}
\includegraphics[scale=0.9]{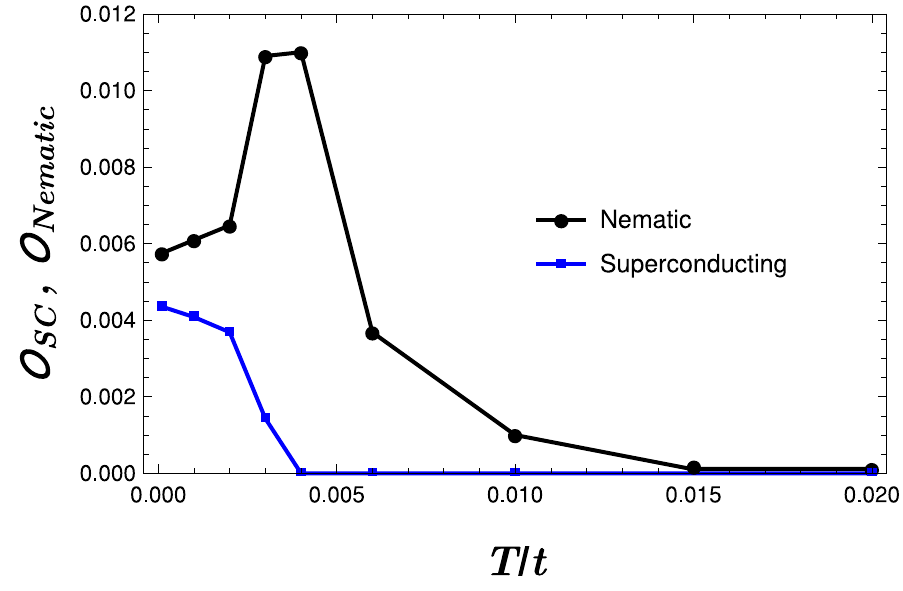}
\end{center}
\vspace{-0.7 cm}
\caption{\label{fig:competition} 
Competition of nematic and superconducting order parameters at doping $n=0.883$, as the temperature is varied.}
\end{figure}

 For the HM as defined in Eqs. (2,3) 
 with $t' = 0.15 t$ and $U = 3 t $  
 \footnote{$t'$ is 
defined with different sign  relative to work \onlinecite{Metzner1}} and all energies measured in units of $t$, the focus is
on the range of dopings around $n = 0.880$,  where a Lifshitz transition from closed to open FS is expected to happen, see Fig.\ref{fig:FSIllustration}.
As a first step, the nematic phase reported in Ref. \cite{Metzner1} is reproduced and the result is shown in Fig.\ref{fig:NematicExample}. 
The competition between the superconducting and 
nematic order parameters  can be traced as a function of the temperature.  The superconducting order is defined 
as an absolute value of $\Delta(\bk)$ 
averaged over the Fermi surface,
$$O_{SC} = \< |\Delta| \>_{FS},$$
while the nematic order as 
\beq
  O_{\rm Nematic} = 2 \int \frac{d^2 \bk}{(2 \pi)^2} (\cos(k_x) - \cos(k_y)) n_F(\tilde \xi,T), 
\eeq 
these definitions are only used to describe the results, while the actual calculations consider the exact FS geometry and a SC gap along it. 
A typical result is presented in 
Fig.\ref{fig:competition}, which shows that nematic order gets notably stronger when superconductivity gets suppressed.  
This suggests that nematic fluctuations are less sensitive to temperature than the superconducting order and hence the nematic order is the 
stronger of the two at non-zero temperature ( $T=0.004 t$ in our example), where the superconductivity becomes strongly suppressed. 

It is necessary to understand at a qualitative level the contributions that drive the nematic transition. 
Consider
\Eq{SigmaEq} for a typical nematic Fermi surface, illustrated in Fig. \ref{fig:NematicExample}(a).  
The first order in $U$ term is irrelevant in the HM as it can be absorbed in the chemical potential.
If we neglect, for the sake of qualitative argument, the superconducting
gap as well as higher frequencies in internal integrations given that low frequencies are the most sensitive to changes of FS shape, then this leads to a simple qualitative formula
\beq \label{simple}
  \Sigma(\bk) \sim U^2 \int d^2 \bq  \frac{\Pi(0,\bq)}{\tilde\xi_{\bk - \bq}},
\eeq 
which suggests that there is a positive contribution to the self-energy at the FS point $\bk$ (meaning, that FS would shrink at this point) 
when it is connected to inside of the (electron-like) FS ($\tilde\xi_{\bk-\bq} < 0 $) by 
the momentum vector $\bq$ at which the 1-loop spin susceptibility, $\chi = -\Pi(\bq)$,  is taken  and a negative contribution otherwise. 
Note that the $1/\tilde\xi$ term is strongly peaked near the FS, but the contributions from the two sides of the FS are of opposite signs and mostly cancel each other if the dispersion near the FS is linear.
However, this is not true near Van Hove singularities, (i.e. in the vicinity of the topological transition), where the dispersion is essentially non-linear, creating a large asymmetry between electrons 
and holes near the FS.  The position of Van Hove singularity relative to the FS determines the dominant contribution:  it is  outside the FS near the  closed corner of the FS (i.e., the Van Hove point is above the Fermi-level) 
(left/right in  Fig.\ref{fig:NematicExample}), leading to the dominance of $1/\tilde \xi >0 $ contribution, while  for the open corner 
(top/bottom in  Fig.\ref{fig:NematicExample}),  the Van Hove spot is inside the FS (i.e. the Van Hove point is below the Fermi level), leading to  dominant $1/\tilde \xi <0$ contribution. 

The fate of the nematic phase is decided by the values of  the self-energy near the corners of the FS, where the sensitivity to self-energy is enhanced by a shallow dispersion, 
so, it is evident that both $\bk$ and $\bk+\bq$  in \eq{simple} should be considered near the corners of the FS.   Therefore, the susceptibilities at momenta $\bq = (0,0)$   and   $\bq = (\pi , \pi)$ (joining 
two different corners) are important and their relative strength  decides the possibility of existence of the nematic phase.
The susceptibility  at $\bq=(0,0)$ gives positive/negative contribution  to the  self-energy at the open/closed corner of the FS receives, not favouring the nematic phase.
On the contrary, 
susceptibility at $\bq = (\pi , \pi)$ (connecting the two corners of the FS), connects the open corner of the FS with, dominantly, $\tilde\chi_{\bk+\bq} >0$  region near the closed 
corner, and, hence, gives
the negative self-energy contribution near the open corner of the FS,  supporting its opening.  
For the studied HM, the susceptibility at the vector connecting the two different corners dominates, 
\beq \label{condition}
-\Pi_{(\pi,\pi)} > -\Pi_{(0,0)},
\eeq
as it is evident in Fig.\ref{fig:NematicExample}(c), 
thus explaining the appearance of the nematic phase. The above considerations highlight the important of using renormalized perturbation
theory, and can be applicable in a more general framework. 
Another example of the same behaviour is offered by the extended Hubbard model, where one adds a nearest-neighbour interaction term: 
$H \to H + V \sum_{<i,j>} n_{i,s} n_{j,s'}$. 
This creates an extra first-order contribution to the self-energy, where the Fermion bubble $\Pi$ is replaced by a tree-level interaction: 
$ U^2 \Pi(\omega,\bq) \to V [ \cos(q_x) + \cos(q_y) ] $.   Our condition (\ref{condition}) then corresponds to $V>0$,  in which case the nematic phase has indeed been 
predicted in Ref. \onlinecite{Vozmediano}. 

In general, similar arguments lead to the physical picture that in the presence of several symmetry-related Van Hove points, with the same energy in the symmetric case,   the symmetry may be broken in a way that some 
Van Hove points are pushed above the Fermi-level, while the others are below the Fermi-level.
This would happen  if the total strength of fluctuations at the wave-vectors connecting the points at the opposite sides of the Fermi-level is larger 
than at the wave-vectors, connecting Van Hove points at the same side of the Fermi-level.     

\begin{figure}
\begin{center}
\includegraphics[scale=0.6]{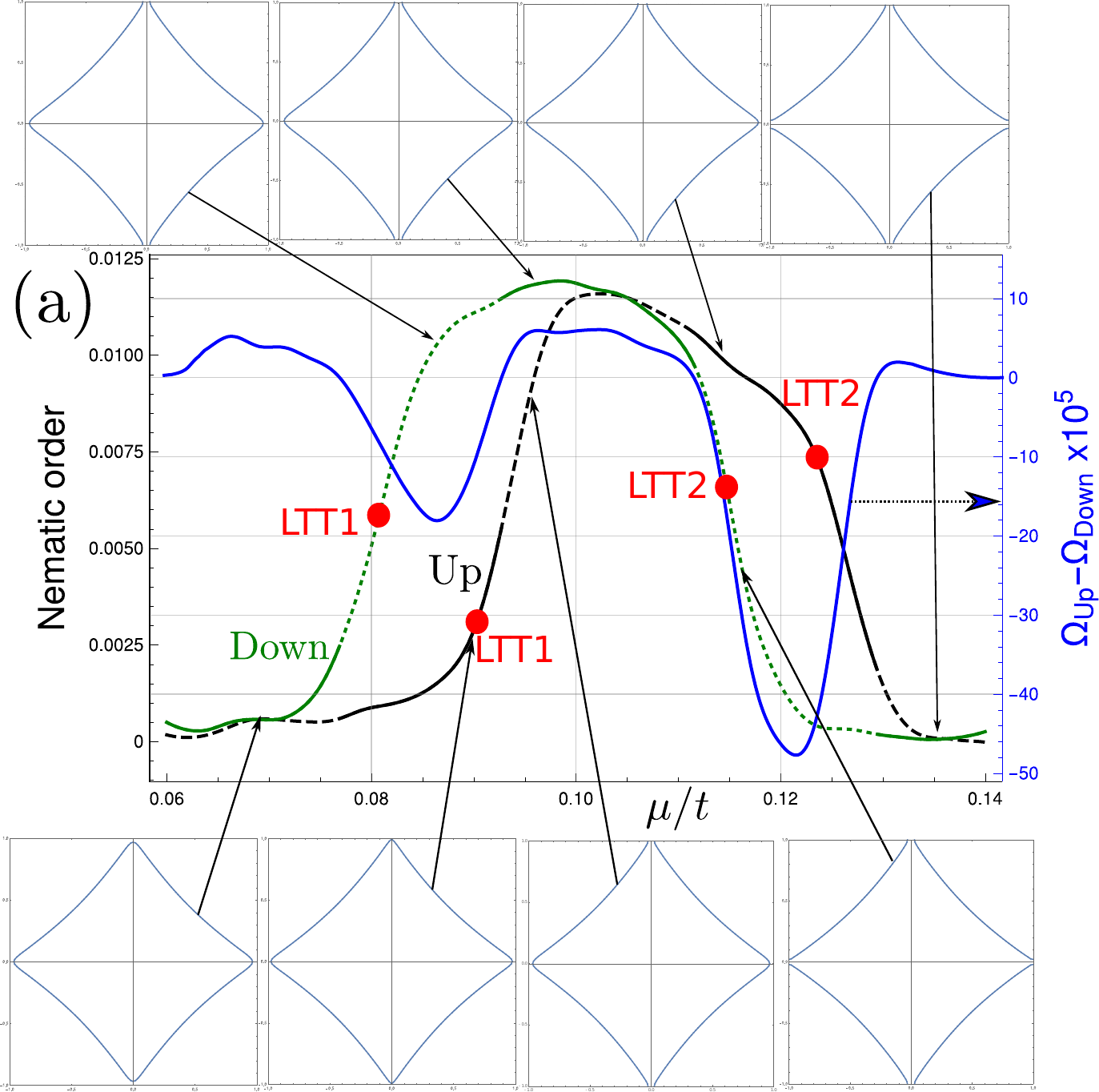}
\includegraphics[scale=0.6]{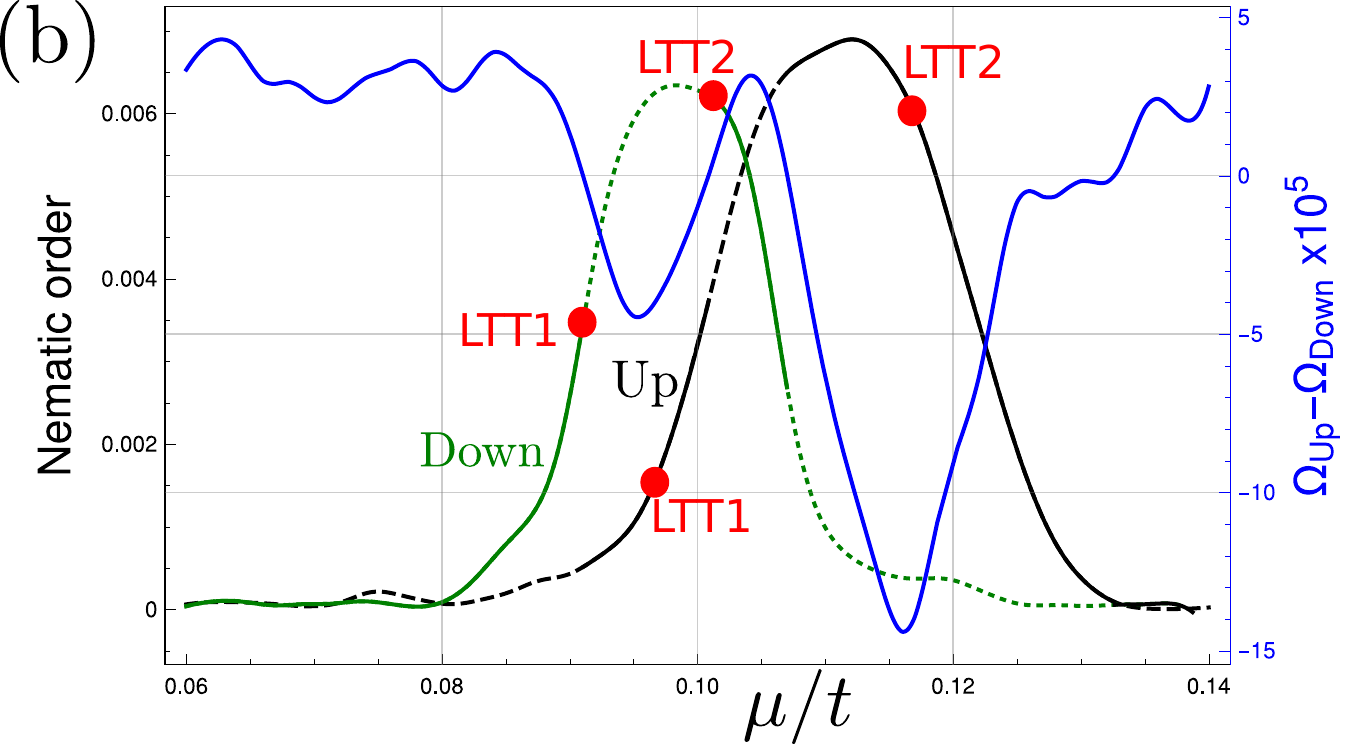}
\end{center}
\vspace{-0.7 cm}
\caption{\label{fig:Omega} 
Hysteresis while evolving up (black) and down (green) in the chemical potential for temperatures  (a) $T=0.004 t$,  (b) $T=0.001 t$. 
The value of the grand canonical potential identifies the preferred state and predicts the points
of first-order phase transitions to and from the nematic phase. The preferred phase is shown with a solid curve, while the metastable phase with a dashed line.  
The 4 red dots indicates the points where the FS undergoes the 
Lifshitz-type topological transition (LTT), which 
are now split into LTT1 (two opposite corners of FS open up)  and LTT2 (two remaining corners open up). Fixed chemical potential corresponds to (almost) fixed filling factor, for example $n(0.08) \approx 0.877 ;\, n(0.10) \approx 0.882;\, n(0.12) \approx 0.888$.}
\end{figure}

\begin{figure}
\begin{center}
\includegraphics[scale=0.3]{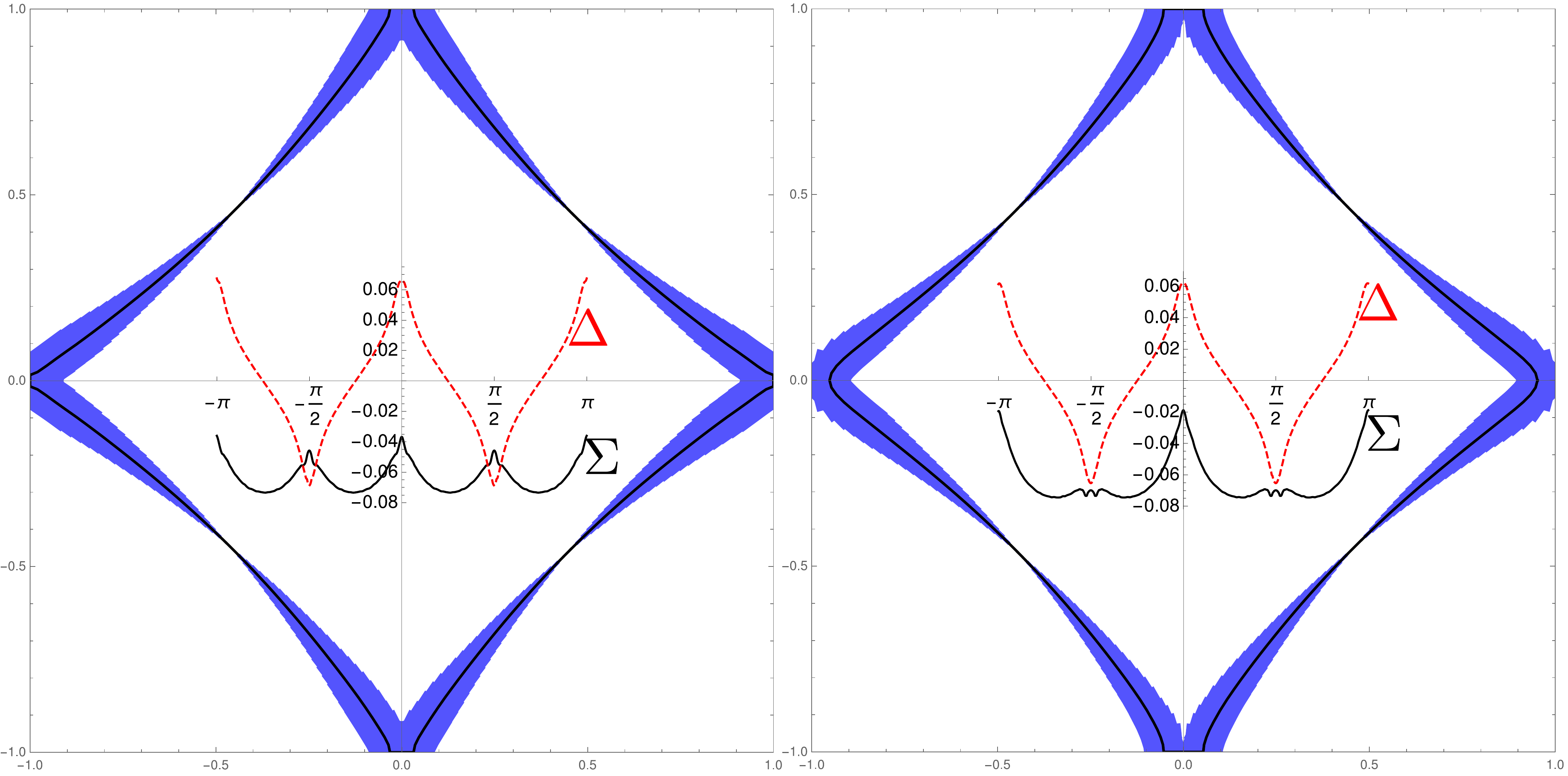}
\end{center}
\vspace{-0.7 cm}
\caption{\label{fig:U=4hyster} 
Illustration of Fermi-surfaces for $U=4 \, t\,$,  $T=0.001\, t$ at the same chemical potential that we get by ``up'' and ``down'' evolution.  The width of the line corresponds to the 
value of superconducting order parameter at the FS: $(width)= \Delta(\vec k)/v_F(\vec k)$ .
Insets shows the self-energy $\Sigma$  and the superconducting gap $\Delta$ at the Fermi-surface  as a  function of angle from $x$-axis.   We see that the two Fermi-surfaces have  different  amplitudes of nematic deformation, 
which illustrates a significant hysteresis. We note that amplitude of $\Delta$ is almost unaffected by nematic order. }
\end{figure}

\begin{figure}
\begin{center}
\includegraphics[scale=0.6]{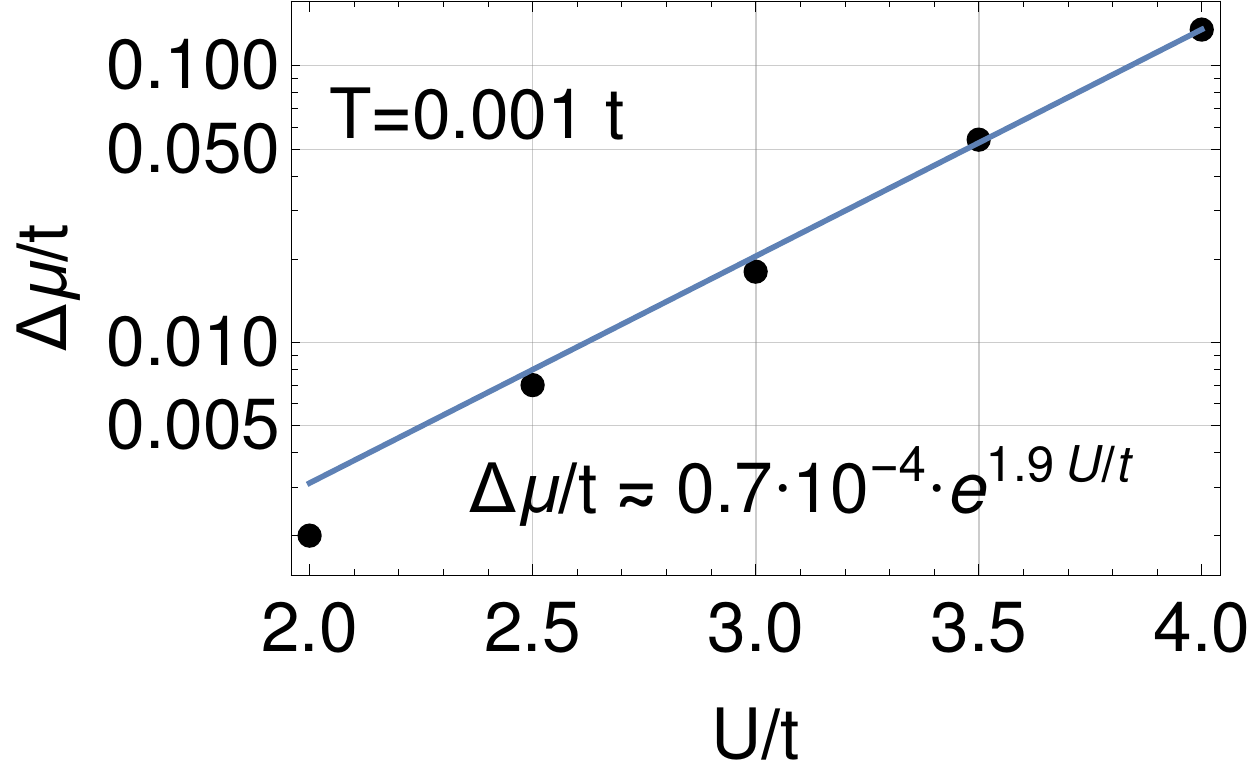}
\end{center}
\vspace{-0.7 cm}
\caption{\label{fig:NematicPlot} 
Width of nematic phase, $\Delta\mu = \mu_\text{max}^\text{``Up''} - \mu_\text{min}^\text{``Down''}$  (measured in terms of chemical potential $\mu' = \mu - U n(\mu)/2$  with subtracted ``trivial'' $U n/2$ term)   
as a function of coupling strength $U$. Temperature is kept fixed to $0.001 t$.  The result is fitted  by an exponential function of $U$:  $\Delta \mu/t \approx 0.7 \cdot 10^{-4}\cdot e^{1.9 U/t} $. }
\end{figure}

 We have found that there may be two competing locally-stable phases near the onset of nematic phase, which poses the question about the exact location of the nematic transition. 
 This question has not been studied before \cite{Metzner1, Vozmediano}, and numerical fluctuations in the implementation of the algorithm   
 chose the phase to which the numerical iterations converged.    To overcome this difficulty, we performed the numerical calculations while smoothly varying the chemical potential,
 allowing to stay
 in the basin of attraction of a particular physical phase.  Smoothly evolving "up'' and  "down" in chemical potential revealed a significant hysteresis  near  the  beginning and the end of the nematic 
 phase, Fig. \ref{fig:Omega}, in particular, the two paths have passed the Lifshitz topological transitions (LTT) at different values of the chemical potential.  
 {  The hysteresis is particularly visible if we increase the coupling,  an example for $U=4$ is shown in Fig.\ref{fig:U=4hyster}.}
 The electron concentrations of the two competing phases at the same chemical potential differ by a tiny amount,  but this difference is important in calculating the difference of free energies.
The results show first order jumps between the phases where  the difference of grand canonical free energy for the coexisting phases changes sign,   Fig. \ref{fig:Omega}. 
This is reflected in the jumps in the nematic order, $O_{\rm Nematic}$, that  are typically  between the small (but non-zero) 
and larger values, suggesting that the nematic order is not a conventional order parameter in this situation and it only reflects on the more complicated Pomeranchuk-type instabilities of the FS.  
{
  From Fig. \ref{fig:Omega} we see that small nematic deformations can appear continuously (without a phase transition),  and that this ``small nematic'' phase is preferred over a ``no-nematic'' phase near the onset
  of nematic transition.  Technically, this allows us to define the ``nematic range'' in chemical potential around the Van Hove singularity as the interval, where either ``up'' or ``down'' numerical evolution in the chemical potential  
produces a nematic phase, and study the width of nematic range around the Van Hove point.  In Fig.\ref{fig:NematicPlot} we find an exponential growth of the width of the nematic phase as the coupling $U$ gets larger
\footnote{Large $U$ is beyond the range of validity of the second-order perturbation theory, but the purpose of using larger values of $U$ in strictly is to reduce the numerical errors in extracting $U$-dependence}.
}

{
\section{Discussion}

Considering the above findings, there are several crucial points to be discussed at this stage. We found that superconductivity and nematicity coexist, as a function of temperature, and the latter is enhanced when the order parameter of the superconducting state gets suppressed. As the FS is partly gapped due to superconductivity, this weakens the nematic state. Conversely, whether superconductivity is affected by the nematic state will take calculations to higher order in the interaction strength to decide.  

Another point to be made is related to the onset of an s-wave superconducting component (d+s) as shown in some recent works \cite{Kaczmarczyk_etal_2016, Kitatani_etal_2017} when the nematic state sets in. This is in agreement with our findings as presented in the inset of Fig.\ref{fig:U=4hyster}, where the amplitude of the superconducting order parameter at $0$ and $\pi/2$ is different. Our method,  allows the presence of all harmonics of the order parameter without restriction to d and s (which are expected to be the dominant ones nevertheless).
Finally, the observation of the first order transition is in agreement with the result that the Lifshitz transition turns to first order in the presence of interactions or magnetic fluctuation \cite{SSJoseph}.  }

To conclude, in this work the self-consistent renormalized perturbation theory is extended to finite temperatures, and the 2D Hubbard model at particular parameters
near a Lifshitz transition is studied.  It is found that the competition of nematic and superconducting orders leads to enhancement of nematic deformations of the FS  at non-zero temperature, where superconductivity
get suppressed. { We provide the conditions that lead to the breaking of the tetragonal $C_4$ symmetry}. The presence of locally stable competing phases in a range of chemical potentials is revealed, with first-order phase transitions between them. { The interval of chemical potentials around the 
Lifshitz transition point  where the nematic phase is observed  is found to grow exponentially with interaction strength $U$. }

\section{Acknowledgements}
We are grateful to Andrey Chubukov, Vladimir Fal'ko, Clifford Hicks, Andrey Katanin, Evgeny Kozik and Jorge Quintanilla for useful discussions.  
S.S. acknowledges financial support from EPSRC through grant EP/l02669X/1 and the Graphene Flagship at the University of Manchester. 
P.R.-L. acknowledges financial support from the US Department of Energy under grant No. DE-FG02-06ER46297, from EPSRC under grant No. EP/H049797/1, project TerMic (Grant No. FIS2014-52486-R, Spanish Government), project CONTRACT (Grant No. FIS2017-83709-R, Spanish Government), and from Juan de la Cierva - Incorporacion program (Ref: I JCI-2015-25315, Spanish Government). JJB's work has been supported by EPSRC through grants EP/H049797/1 and EP/P002811/1.

\section{APPENDICES}

\subsection{A. Extra term in the Nambu form of the Hamiltonian}
We start from the initial tight binding Hamiltonian $H_{0}$, where $\xi_{\textbf{k}} = \epsilon_{\textbf{k}} - \mu$:
\begin{eqnarray}
H_{0} & = & \sum_{\textbf{k},\sigma}c_{\textbf{k}}^{\dagger}\xi_{\textbf{k}}c_{\textbf{k}}
 = \sum_{\textbf{k}}\left(\begin{array}{cc}
a_{\textbf{k},\uparrow}^{\dagger} & a_{-\textbf{k},\downarrow}^{\dagger}
\end{array}\right)\left(\begin{array}{cc}
\xi_{\textbf{k}} & 0\\
0 & \xi_{-\textbf{k}}
\end{array}\right)\left(\begin{array}{c}
a_{\textbf{k},\uparrow}\\
a_{-\textbf{k},\downarrow}
\end{array}\right)\\
& = & \sum_{\textbf{k}}a_{\textbf{k},\uparrow}^{\dagger}\xi_{\textbf{k}}a_{\textbf{k},\uparrow} + 
a_{-\textbf{k},\downarrow}^{\dagger}\xi_{-\textbf{k}}a_{-\textbf{k},\downarrow}
\end{eqnarray}
On the other side, the Nambu Hamiltonian $H_{N}$, for zero superconducting gap $(\Delta_{\textbf{k}}=0)$ is defined as
\begin{eqnarray}
H_{N}(\Delta_{\textbf{k}}=0) & = & \sum_{\textbf{k},\sigma}\psi_{\textbf{k}}^{\dagger}\xi_{\textbf{k}}\sigma_{3}\psi_{\textbf{k}}
 = \sum_{\textbf{k}}\left(\begin{array}{cc}
a_{\textbf{k},\uparrow}^{\dagger} & a_{-\textbf{k},\downarrow}
\end{array}\right)\left(\begin{array}{cc}
\xi_{\textbf{k}} & 0\\
0 & -\xi_{-\textbf{k}}
\end{array}\right)\left(\begin{array}{c}
a_{\textbf{k},\uparrow}\\
a_{-\textbf{k},\downarrow}^{\dagger}
\end{array}\right)\\
& = & \sum_{\textbf{k}}a_{\textbf{k},\uparrow}^{\dagger}\xi_{\textbf{k}}a_{\textbf{k},\uparrow} - 
a_{-\textbf{k},\downarrow}\xi_{-\textbf{k}}a_{-\textbf{k},\downarrow}^{\dagger}
\end{eqnarray}
For fermionic fields, since the anticommutator relation $\{a_{\textbf{k},\sigma}^{\dagger}, a_{\textbf{q},\sigma'}\} = \delta(\textbf{k} - \textbf{q})\delta_{\sigma,\sigma'}$, holds, therefore,
when, applying this relation to $H_{N}$, we find
\begin{eqnarray}
\nonumber
H_{N}(\Delta_{\textbf{k}}=0)
& = & \sum_{\textbf{k}}\xi_{\textbf{k}}a_{\textbf{k},\uparrow}^{\dagger}a_{\textbf{k},\uparrow} - 
\xi_{-\textbf{k}}a_{-\textbf{k},\downarrow}a_{-\textbf{k},\downarrow}^{\dagger}\\
\nonumber
& = & \sum_{\textbf{k}}\xi_{\textbf{k}}a_{\textbf{k},\uparrow}^{\dagger}a_{\textbf{k},\uparrow} - 
\xi_{-\textbf{k}}\left[ 1 - a_{-\textbf{k},\downarrow}a_{-\textbf{k},\downarrow}^{\dagger}
\right]\\
\nonumber
& = & \sum_{\textbf{k}}\left[a_{\textbf{k},\uparrow}^{\dagger}\xi_{\textbf{k}}a_{\textbf{k},\uparrow} + 
 a_{-\textbf{k},\downarrow}\xi_{-\textbf{k}}a_{-\textbf{k},\downarrow}^{\dagger}\right] - \sum_{\textbf{k}}\xi_{-\textbf{k}}\\
& = & H_{0} - \sum_{\textbf{k}}\xi_{-\textbf{k}}
\end{eqnarray}
Therefore
\begin{equation}
H_{0} = H_{N}(\Delta_{\textbf{k}}=0) + \sum_{\textbf{k}}\xi_{-\textbf{k}}
\end{equation}
In our particular model, with symmetry respected under momentum inversion $\xi_{-\textbf{k}} = \xi_{\textbf{k}}$, then
\begin{equation}
H_{0} = H_{N}(\Delta_{\textbf{k}}=0) + \sum_{\textbf{k}}\xi_{\textbf{k}}
\end{equation}
which provides the extra term in the energy of the system.

\subsection{B. Tree level of grang canonical free energy}
In this section, we obtain the grand canonical free energy $\Omega_{0}$ for a Gaussian fermionic field with a constant term.
As we have seen the normal-ordered Hamiltonian written in the Nambu formalism contains an additional constant term that should enter into the final expression of the free energy.
\begin{equation}
H_{0} = H_{N}(\Delta_{\textbf{k}}=0) + \sum_{\textbf{k}}\xi_{-\textbf{k}}
\end{equation}
Using the fact that for fermionic fields, the anticommutation relation $\{a_{\textbf{k}}^{\dagger}, a_{\textbf{q}}\} = \delta_{\textbf{k},\textbf{q}}$ holds, 
%
then the fermionic Hamiltonian is written as $H = \int_{k}H_{\textbf{k}}$
\begin{eqnarray}
H_{\textbf{k}} & = & \xi_{-\textbf{k}} + \frac{\hbar\omega_{\textbf{k}}}{2}\left[ a_{\textbf{k}}^{\dagger}a_{\textbf{k}} - a_{\textbf{k}}a_{\textbf{k}}^{\dagger} \right]\\
& = & \xi_{-\textbf{k}} + \hbar\omega_{\textbf{k}}\left[ a_{\textbf{k}}^{\dagger}a_{\textbf{k}} - \frac{1}{2} \right].
\end{eqnarray}
In particular,
\begin{itemize}
\item For the superconducing state, we have $\hbar\omega_{\textbf{k}}(\Delta) = \pm E_{\textbf{k}} = \pm\sqrt{ \xi_{\textbf{k}} + \abs{\Delta_{\textbf{k}}}^{2} }$
\item For the normal state, we have $\hbar\omega_{\textbf{k}} = \pm \xi_{\textbf{k}} = \lim_{\Delta\to 0}\hbar\omega_{\textbf{k}}(\Delta)$
\end{itemize}
The fermionic partition function is defined as
\[\mathcal{Z} = \Tr{e^{-\beta\hat{H}}} = \prod_{\textbf{k}}\mathcal{Z}_{\textbf{k}} = \prod_{\textbf{k}}\Tr{e^{-\beta\hat{H}_{\textbf{k}}}}\]
Due to Pauli exclusion principle $\left(\ket{n} = 0 \,\,\,\forall n>1\right)$, this trace is easily evaluated: $\{\ket{0}, \ket{1}\}$, then
\[\mathcal{Z}_{\textbf{k}} = \Tr{e^{-\beta\hat{H}_{\textbf{k}}}} = \bra{0}e^{-\beta\hat{H}_{\textbf{k}}}\ket{0} + \bra{1}e^{-\beta\hat{H}_{\textbf{k}}}\ket{1}\]
Each term is evaluated separately, taking into account that $a_{\textbf{k}}\ket{0} = 0$, $a_{\textbf{k}}\ket{1} = \ket{0}$, and $a_{\textbf{k}}a_{\textbf{k}}^{\dagger}\ket{0} = a_{\textbf{k}}\ket{1} = \ket{0}$.
%

Then we have
\begin{eqnarray}
\bra{0}e^{-\beta\hat{H}_{\textbf{k}}}\ket{0}
& = & \bra{0}e^{-\beta\xi_{-\textbf{k}}}e^{-\beta\hbar\omega_{\textbf{k}}a_{\textbf{k}}^{\dagger}a_{\textbf{k}}}e^{\beta\frac{\hbar\omega_{\textbf{k}}}{2}}\ket{0}\nonumber\\
& = & e^{-\beta\xi_{-\textbf{k}}}e^{\beta\frac{\hbar\omega_{\textbf{k}}}{2}}
\end{eqnarray}

\begin{eqnarray}
\bra{1}e^{-\beta\hat{H}_{\textbf{k}}}\ket{1}
& = & \bra{1}e^{-\beta\xi_{-\textbf{k}}}e^{-\beta\hbar\omega_{\textbf{k}}a_{\textbf{k}}^{\dagger}a_{\textbf{k}}}e^{\beta\frac{\hbar\omega_{\textbf{k}}}{2}}\ket{1}\nonumber\\
& = & e^{-\beta\xi_{-\textbf{k}}}e^{-\beta\frac{\hbar\omega_{\textbf{k}}}{2}}
\end{eqnarray}
Therefore
\begin{eqnarray}
\mathcal{Z}_{\textbf{k}}
& = & \bra{0}e^{-\beta\hat{H}_{\textbf{k}}}\ket{0} + \bra{1}e^{-\beta\hat{H}_{\textbf{k}}}\ket{1}\nonumber\\
& = & e^{-\beta\xi_{-\textbf{k}}}\left[ e^{\beta\frac{\hbar\omega_{\textbf{k}}}{2}} + e^{-\beta\frac{\hbar\omega_{\textbf{k}}}{2}} \right]
\end{eqnarray}
The grand canonical free energy can be computed as
\begin{eqnarray}
\Omega_0
& = & - T\log\left(\mathcal{Z}\right)\nonumber = - T\int\frac{d^{2}\textbf{k}}{(2\pi)^{2}}\log\left(\mathcal{Z}_{\textbf{k}}\right)\nonumber\\
& = &  - T\int\frac{d^{2}\textbf{k}}{(2\pi)^{2}}\left[ \log\left(e^{-\beta\xi_{-\textbf{k}}}\right) + \log\left(e^{\beta\frac{\hbar\omega_{\textbf{k}}}{2}} + e^{-\beta\frac{\hbar\omega_{\textbf{k}}}{2}}\right)\right]\nonumber\\
& = &  \int\frac{d^{2}\textbf{k}}{(2\pi)^{2}}\left[ \xi_{-\textbf{k}} - \frac{\hbar\omega_{\textbf{k}}}{2} - T\log\left( 1 + e^{-\beta\hbar\omega_{\textbf{k}}}\right) \right]
\end{eqnarray}

By summing the contribution of the two branches, $\hbar\omega_{\textbf{k}}(\Delta) = \lambda E_{\textbf{k}}$ or $\hbar\omega_{\textbf{k}} = \lambda \xi_{\textbf{k}}$ for $(\Delta_{\textbf{k}} = 0)$ with $\lambda = \pm 1$, then
\begin{eqnarray}
\Omega_0
& = &  \int\frac{d^{2}\textbf{k}}{(2\pi)^{2}}\left[ \xi_{-\textbf{k}} - \sum_{\lambda=\pm 1}\left( \frac{\lambda E_{\textbf{k}}}{2} + T\log\left( 1 + e^{-\beta\lambda E_{\textbf{k}}}\right) \right) \right]\nonumber\\
& = &  \int\frac{d^{2}\textbf{k}}{(2\pi)^{2}}\left[ \xi_{-\textbf{k}} - T\left( \beta E_{\textbf{k}} + 2\log\left( 1 + e^{-\beta E_{\textbf{k}}}\right) \right) \right]\nonumber\\
& = &  - \int\frac{d^{2}\textbf{k}}{(2\pi)^{2}}\left[ \left( E_{\textbf{k}} - \xi_{-\textbf{k}} \right) + 2T\log\left( 1 + e^{-\beta E_{\textbf{k}}}\right)  \right],
\end{eqnarray}
which is Eq. (21) in the main text.




\begin{spacing}{0}
\setlength{\bibsep}{1pt}
\bibliographystyle{hplain}
  \bibliography{References}
\end{spacing}
\end{document}